\documentstyle[prb,twocolumn,aps,psfig,amssymb,floats]{revtex}

\begin{document}
\draft

\twocolumn[\hsize\textwidth\columnwidth\hsize\csname @twocolumnfalse\endcsname


\title{The Kondo lattice model with correlated conduction electrons}
\author{Tom Schork and Stefan Blawid}
\address{Max-Planck-Institut f\"ur Physik komplexer Systeme,
N\"othnitzer Str.\ 38, 01187 Dresden, Germany}
\author{Jun-ichi Igarashi}
\address{Faculty of Engineering, Gunma University, Kiryu, Gunma 376, Japan}

\maketitle


\begin{abstract}
  We investigate a Kondo lattice model with correlated conduction
  electrons.  Within dynamical mean-field theory the model maps onto
  an impurity model where the host has to be determined
  self-consistently.  This impurity model can be derived from an
  Anderson-Hubbard model both by equating the low-energy excitations
  of the impurity and by a canonical transformation. On the level of
  dynamical mean-field theory this establishes the connection of the two
  lattice models.  The impurity model is studied numerically by an
  extension of the non-crossing approximation to a two-orbital
  impurity.  We find that with decreasing temperature the conduction
  electrons first form quasiparticles unaffected by the presence of
  the lattice of localized spins. Then, reducing the temperature
  further, the particle-hole symmetric model turns into an insulator.
  The quasiparticle peak in the one-particle spectral density splits
  and a gap opens. The size of the gap increases when the correlations
  of the conduction electrons become stronger. These findings are
  similar to the behavior of the Anderson-Hubbard model within
  dynamical mean-field theory and are obtained with much less
  numerical effort.
\end{abstract}



\vskip2pc]

\section{Introduction}
\label{sec:intro}

The usual explanation for the formation of heavy fermions in compounds with
rare-earth or actinide elements is based on the Kondo
effect.\cite{Fulde88,HewsonBook} Thereby, the characteristic low-energy scale
arises from the spin-screening of the local moments by a non-interacting
electron gas. The periodic Anderson model, as well as the Kondo lattice model,
are considered as the most promising candidates to at least qualitatively
describe the rich physics of these materials. However, these models
fail\cite{Fulde93} to describe correctly the temperature scale of the
heavy-fermion behavior in the electron-doped cuprate $\rm Nd_{2-x}Ce_xCuO_4$
discovered a few years ago.\cite{Brugger93} It has therefore been suggested
that the strong correlations among the doped carriers which are usually
neglected in the Kondo effect play a crucial role and may enhance the Kondo
temperature significantly.\cite{Fulde93} Subsequently, efforts have been
undertaken in order to study the influence of the correlations among the
conduction electrons on the Kondo effect, both for
impurity\cite{Furusaki94,Li95,Froejdh96,Schork94,Khaliullin95,Igarashi95,%
Igarashi95b,Schork96} and lattice\cite{Shibata96,Itai96,Schork97b}
models.

These correlations are usually introduced by adding a Hubbard-type
interaction among the conduction electrons to the Anderson model. In
Ref.~\onlinecite{Schork97b} the resulting (periodic) Anderson-Hubbard
model was investigated within the dynamical mean-field
theory.\cite{Georges96,Pruschke95} It was shown that it maps onto an
impurity model where the impurity consists of a unit cell of the
lattice model, i.e., of two correlated orbitals.  This model still
includes charge fluctuations on the $f$ orbital which should not
influence the low-energy behavior in the Kondo limit. It would be
desirable to eliminate these degrees of freedom as this reduces the
number of states of the impurity and thus reduces the computational
effort to investigate the impurity model. Hence, we are aiming at a
Kondo-Hubbard model as the counterpart of the Anderson-Hubbard model
just as in the case of uncorrelated conduction
electrons.\cite{Schrieffer66,Lacroix79,Proetto81}

In the next section, we introduce the Kondo-Hubbard model and map it
to an impurity model using the dynamical mean-field approximation. In
Sec.~\ref{sec:mapping} we relate the derived effective impurity model
with the one of the Anderson-Hubbard model. We have applied the
resolvent perturbation theory\cite{Keiter84} to investigate the
effective impurity model for the Kondo-Hubbard model.  We present some
results in Sec.~\ref{sec:results} and finally conclude in
Sec.~\ref{sec:conclusion}.

\section{The Kondo-Hubbard model}
\label{sec:models}

The (periodic) Kondo-Hubbard model is given by
\begin{eqnarray}
H_{\rm K} & = & H_c + H_S ~,
\label{K_hamil} \\
H_c & = & \sum_{k\sigma} 
\left( \epsilon_k-\frac{1}{2} U_c \right) 
c^\dagger_{k\sigma} c^{\phantom{\dagger}}_{k\sigma} 
+ U_c \sum_i n^c_{i\uparrow} n^c_{i\downarrow} ~,
\label{hamil_c} \\
H_S & = & J \sum_i \vec{S}^f_i \vec{S}^c_i ~.
\label{hamil_S}
\end{eqnarray}
Here, the $c$ operators refer to the conduction electrons which are
described by a Hubbard model with on-site repulsion
$U_c$. $\vec{S}^{c(f)}_i$ denotes the spin of an electron at site $i$
in the conduction band orbital $c$ and the impurity orbital $f$,
respectively. For simplicity we consider in this paper the
particle-hole symmetric model only, which implies $\mu =0$.

We closely follow the treatment of the usual Kondo lattice model in infinite
dimensions given in Ref.~\onlinecite{Matsumoto95}. We introduce Fermion
operators $d_{i\sigma}$ to represent the $f$ spin
\begin{equation}
\vec{S}^f_i = \frac{1}{2} \sum_{\sigma\sigma'} d^\dagger_{i\sigma}
\vec{\sigma}_{\sigma\sigma'} d^{\phantom{\dagger}}_{i\sigma'}
\end{equation}
with local constraints
\begin{equation}
Q_i = \sum_\sigma d^\dagger_{i\sigma} d^{\phantom{\dagger}}_{i\sigma} = 1
\end{equation}  
for all $i$. To enforce the constraints we add
\begin{equation}
H_U = \frac{1}{2}U\sum_i (Q_i-1)^2 
\end{equation} 
to the Hamiltonian and take the limit $U\to\infty$.\cite{Matsumoto95} The
Hamiltonian $H_{\rm K}$ includes three two-particle parts proportional
to $J$, $U_c$ and $U$, respectively. Given the noninteracting Green's
function, $G_0(\vec{k},\omega)$, of the Kondo-Hubbard model 
\begin{equation}
G_0^{-1}(k,\omega) = \left( 
\begin{array}{cc}
\omega+\frac{1}{2} U_c-\epsilon_k & 0 \\
0 & \omega+\frac{1}{2} U
\end{array}
\right)
\end{equation} 
the self-energies are defined by Dyson's equation  
\begin{equation}
G(k,\omega) = \left( G_0^{-1}(k,\omega) - 
\Sigma(k,\omega) \right)^{-1}.
\end{equation}
The dynamical mean-field theory assumes a momentum-independent
self-energy matrix\cite{Ohkawa91} $\Sigma(k,\omega) \rightarrow
\Sigma(\omega)$ which holds in the limit of infinte
dimensions.\cite{MuellerHartmann89a} Following the arguments given in
Ref.\onlinecite{Schork97b} we can calculate $\Sigma(\omega)$ from the
impurity model
\begin{eqnarray}
H_{\rm K,imp} & = & H_{\rm loc} + H_{\rm med} ~,
\label{K_hamil_imp}
\\
H_{\rm loc} & = & 
-\frac{1}{2} U_c \sum_\sigma c^\dagger_{\sigma} c^{\phantom{\dagger}}_{\sigma}
+ U_c n^c_\uparrow n^c_\downarrow 
+ J \vec{S}^f \vec{S}^c
\label{K_hamil_loc}
\\
H_{\rm med} & = & 
\sum_{k\sigma} E_k \alpha^\dagger_{k\sigma}\alpha^{\phantom{\dagger}}_{k\sigma}
+ \sum_{k\sigma} \left( \Gamma_k c^\dagger_\sigma
\alpha^{\phantom{\dagger}}_{k\sigma} + {\rm H.c.} \right) ~.
\label{K_hamil_med}
\end{eqnarray}
with the self-consistency equation
\begin{equation}
\label{eq:selfcon}
\Delta(\omega) = \sum_{\vec{k}} \frac{|\Gamma_k|^2}
{\omega-E_k} = \frac{W^2}{4} \, {\cal G}_c(\omega). 
\end{equation}
Equation (\ref{eq:selfcon}) is valid for a semi-elliptic density of
states of the conduction electrons
\begin{equation}
\rho_0(\epsilon) = \frac{2}{\pi\,W^2} \,\sqrt{W^2-\epsilon^2}\; . 
\end{equation}
In all our calculations we choose $W=1$ as unit of energy.

In order to solve numerically the impurity model~(\ref{K_hamil_imp})
we make use of the resolvent perturbation theory introduced in
Ref.~\onlinecite{Keiter84}. It serves as a basis for the non-crossing
approximation to the Anderson impurity model\cite{Bickers87} which has
been successfully extended to investigate the finite-$U$ Anderson
impurity model,\cite{Pruschke89} the Emery model in the dynamical mean
field theory,\cite{Lombardo96} and the Anderson-Hubbard model in this
approximation.\cite{Schork97b} As described in
Ref.\onlinecite{Schork97b}, the occuring coupled integral equations
are solved numerically by introducing defect
propagators\cite{Bickers87b} and making use of the fast Fourier
transformation.\cite{Lombardo96} 

\section{Relation with the Anderson-Hubbard model}
\label{sec:mapping}

It is well known that the Anderson and Kondo impurity model are
mutually related,\cite{Schrieffer66} as well as the lattice
models.\cite{Lacroix79,Proetto81} We expect that also the lattice
models which include interactions among the conduction electrons show
the same low-energy behavior. Within the dynamical mean-field theory
this should be true provided the excitation energies of the
corresponding impurities are the same in both models. In
Ref.\onlinecite{Schork97b} we have derived the effective impurity
model $H_{\rm A,imp}=H_{\rm med}+H_{\rm loc}$ for the Anderson-Hubbard
model. The local part reads
\begin{eqnarray}
\label{eq:imp}
H_{\rm loc}  & = &  
-\frac{1}{2} U_c \sum_\sigma c^\dagger_{\sigma} c^{\phantom{\dagger}}_{\sigma}
+ U_c n^c_\uparrow n^c_\downarrow 
+ V \sum_\sigma \left( c^\dagger_\sigma f^{\phantom{\dagger}}_\sigma
+ {\rm H.c.} \right) \nonumber \\
 & & - \frac{1}{2} U_f
\sum_\sigma f^\dagger_{\sigma} f^{\phantom{\dagger}}_{\sigma}
+ U_f n^f_\uparrow n^f_\downarrow ~.
\end{eqnarray}
Note that the mapping being discussed here is based on the impurity
models and not on the original lattice models. Therefore, the same
restrictions apply as to the dynamical mean-field theory or the limit of
infinite coordination number of the underlying lattice. In particular, spatial
correlations are not included, so that an RKKY interaction will not
occur,\cite{Jarrell95} in contrast to the treatment in finite
dimensions.\cite{Proetto81}

It has been found in Ref.~\onlinecite{Schork97b} that two characteristic
features occur at low energies for the Anderson-Hubbard model. At intermediate
temperatures, a quasiparticle peak is formed at the chemical potential which
corresponds to the quasiparticle peak observed for a pure Hubbard
model:\cite{Ohkawa91,Jarrell92,Georges92} In this temperature regime the
presence of the $f$ orbitals does not influence the low-energy behavior. When
lowering the temperature a gap opens. As in the case of free conduction
electrons, this is interpreted as a hybridization gap which arises from the
level crossing of the quasiparticle band with dynamically created states at
the chemical potential (Kondo-resonance states).\cite{Yamada85,Jarrell95}

\begin{table}
\begin{center}
\begin{tabular}{c|c|c}
& AHM &  KHM \\
\hline
$E(f^1 c^0)$ 
& $ \displaystyle -\frac{U_f+U_c}{4} - \frac{1}{4}\sqrt{(U_f-U_c)^2+16V^2}$ 
& 0 
\\
$E(f^1 c^1 (S))$ 
& $ \displaystyle -\frac{U_f+U_c}{4}-\frac{1}{4}\sqrt{(U_f+U_c)^2+64V^2}$ 
& $\displaystyle - \frac{U_c}{2} - \frac{3J}{4}$ 
\\
$E(f^1 c^1 (T))$ 
& $\displaystyle -\frac{U_f+U_c}{2}$ 
& $\displaystyle -\frac{U_c}{2} + \frac{J}{4}$
\\
$E(f^1 c^2)$ 
& $ \displaystyle -\frac{U_f+U_c}{4} + \frac{1}{4}\sqrt{(U_f-U_c)^2+16V^2}$ 
& 0 
\\
\end{tabular}
\vspace*{0.5cm}
\caption{Eigenenergies of the impurity, $H_{\rm loc}$, in the Anderson- (AHM)
and Kondo-Hubbard model (KHM) in the symmetric case. The indicated occupation
is the one for $V\to 0$. $S(T)$ stands for singlett (triplett) configuration,
resp. States not included for the AHM have zero $f$ occupation in the limit
$V\to 0$ and no counterpart in the KHM.}
\label{tab:energies}
\end{center}
\end{table}
Both features are related to transition energies of the impurity which
generate a Kondo effect on a certain energy scale when coupled to a
bath. The eigenenergies of the impurity (\ref{eq:imp}) where the $f$
orbital is singly occupied when $V\to 0$ are listed in
Tab.~\ref{tab:energies}.  The Hubbard-like quasiparticle peak is not
influenced by the presence of the $f$ orbitals and results therefore
from the $c$ charge degree of freedom. The relevant transition in
$H_{\rm loc}$ is $f^1 c^0 \to f^1 c^1 (S)$ ($S$ stands for singlet)
with its energy given by $\Delta E_1 = E(f^1c^0) - E(f^1c^1(S))$. The
opening of the gap is found at lower temperatures and it is related to
the singlet-triplet excitation $f^1 c^1(T) \to f^1 c^1(S)$ (the lowest
transition energy of the impurity) with energy $\Delta E_2 =
E(f^1c^1(T)) - E(f^1c^1(S))$.

The two impurity models generate the same low-energy behavior,
provided their corresponding low-energy transitions have identical
energies. The eigenenergies of the Kondo impurity (\ref{K_hamil_loc})
are listed in Tab.~\ref{tab:energies} as well. We determine the
parameters of the Kondo-Hubbard model by equating the corresponding
transition energies. The exchange coupling constant $J$ is obtained
from $E(f^1 c^1(T)) - E(f^1 c^1(S)) = \Delta E_2 = J$
\begin{eqnarray}
J & = & \frac{1}{4} \sqrt{(U_f+U_c)^2 + 64 V^2} - \frac{1}{4}
(U_f+U_c) \nonumber \\
& = & \frac{8V^2}{U_f+U_c} + o(V^4) ~.
\label{J}
\end{eqnarray}
The Hubbard interaction in the Kondo-Hubbard model, which we denote by $\tilde
U_c$ in order to distinguish it from the Anderson-Hubbard model, is obtained
from $E(f^1 c^0) - E(f^1 c^1(S)) = \Delta E_1 = \tilde U_c /2 + 3/4 J$
\begin{eqnarray}
\tilde U_c  & = &  \frac{3}{8} (U_f+U_c) 
- \frac{1}{2} \sqrt{(U_f-U_c)^2 + 16 V^2} \nonumber \\
& & + \frac{1}{8} \sqrt{(U_f+U_c)^2 + 64 V^2}
\nonumber \\
& = & U_c \left( 1 -
\frac{8V^2}{U_f^2-U_c^2} \right) + o(V^4) ~.
\label{Uc}
\end{eqnarray}

Whereas the appearance of the exchange coupling $J$ as well as its magnitude
for small $V$ is expected (the intermediate state requires a doubly occupied
$c$ orbital with energy $U_c/2$), the modification of the local interaction of
the conduction electrons might be surprising. However, if $f$ charge
fluctuations are possible, a doubly occupied $c$ orbital may lower its energy
by virtual transitions onto the $f$ orbital. To compensate, the effective
repulsion is lowered when the $f$ occupancy is fixed.

The same relations between the parameters of the Anderson- and Kondo-Hubbard
model are obtained by performing a canonical transformation of the molecule
where one eliminates the charge degrees of freedom of the $f$ orbital in
perturbation theory in the hybridization $V$. This is shown in
App.~\ref{app:can}.

\section{Results}
\label{sec:results}

\begin{figure}
\centerline{\psfig{file=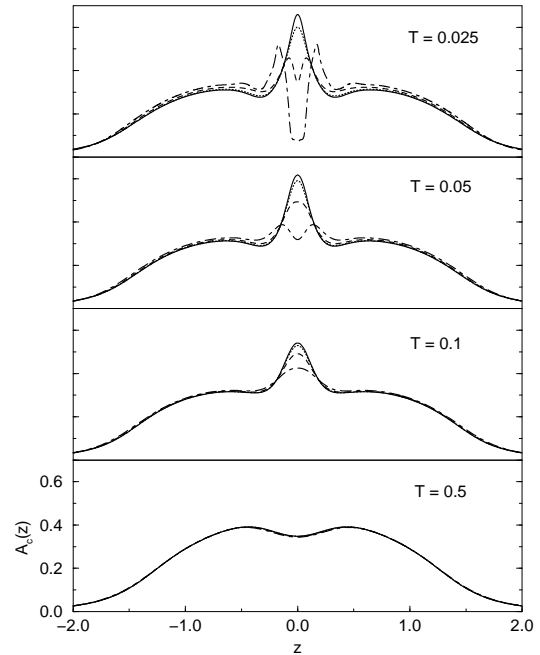,width=8cm}}
\caption{Local density of states $A_c(z)$ in the Kondo-Hubbard model for $U_c
= 1$ and $J = 0$ (solid), $0.05$ (dotted), $0.1$ (dashed), and $0.15$
(dot-dashed line) at different temperatures.}
\label{fig:spectra}
\end{figure}
In Fig.~\ref{fig:spectra} we display the local spectral functions of
the conduction electrons which were obtained by the treatment outlined
in Sec.~\ref{sec:models} for $U_c = 1$ and various $J$ at different
temperatures $T$. The results found are typical for the Kondo-Hubbard
model. At comparatively high temperatures ($T>0.5$), only two broad
maxima are seen which correspond to upper and lower Hubbard band of
the conduction electrons. When lowering the temperature, a peak
emerges at the chemical potential. The spectra at these temperatures
are almost independent of the coupling $J$ to the $f$ spin for small
$J \lesssim 0.1$, and coincide with the spectra calculated for the
pure Hubbard model ($J=0$). Therefore, we attribute this peak to the
quasiparticle peak of the Hubbard
model\cite{Jarrell92,Georges92,Ohkawa92} which describes the
conduction electrons. At these moderate temperatures the $c$ and $f$
electrons are almost decoupled. The quasiparticle peak corresponds to
the $c^1 \to c^0$ transition in the $c$-$f$ molecule. Coupled to the
bath, this transition gives rise to a Kondo effect, which sets a
characteristic energy scale. Assuming $J=0$ and a structureless bath
with $\Delta(0)=2/\pi$, the impurity model reduces to the
symmetric Anderson model and the Kondo temperature is estimated
to $T_K\sim 0.5$.\cite{Tsvelick83} (This rather large scale is due to
the large coupling of the impurity to the bath which is of the order
of the band-width $D$.)  Notice that the given value of $T_K$
provides only a rough estimation of the corresponding energy scale in
our calculation. First of all, the assumption of a structureless
$\Delta$ is not valid. Moreover, neglecting vertex corrections in the
NCA for the symmetric Anderson model, as we have done, underestimates
the Kondo temperature.\cite{Pruschke89}

When we reduce the temperature further, we observe that the
quasiparticle peak gets split (Figs.~\ref{fig:spectra}). This
splitting results from the influence of the $f$ spins since its
characteristic temperature, as well as its magnitude depend on
$J$. Therefore, the $f$ and $c$ subsystems are no longer decoupled in
this temperature regime. At the lowest temperatures that we reach
numerically, and which are lower when $J$ becomes smaller, a gap
begins to open as is expected for the symmetric model. Qualitatively,
these findings fit to the scenario encountered in the Anderson model
with free conduction electrons, where below a characteristic
temperature a local Kondo effect takes place, in which the conduction
electrons screen the $f$ spin, which leads to a resonance at the
chemical potential in the particle-hole symmetric case. These
dynamically generated states cross the conduction-band states, one
finds a splitting of the conduction band with a gap at the chemical
potential, and the system becomes an
insulator.\cite{Yamada85,Jarrell95,Georges96}

\begin{figure}
\centerline{\psfig{file=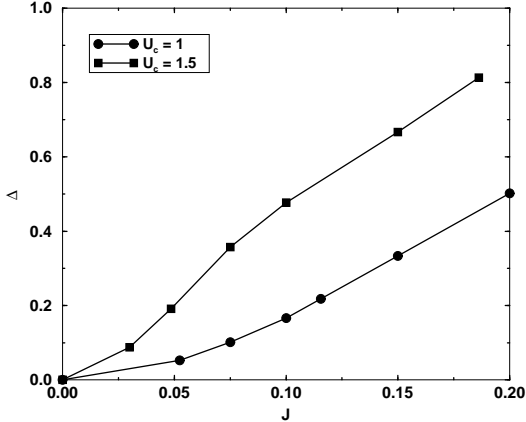,width=8cm,angle=270}}
\caption{Gaps of the Kondo-Hubbard model for $U_c = 1.0$ and $1.5$. The lines
are guides to the eye.}
\label{fig:gaps}
\end{figure}
Quantitatively, however, these effects depend not only on the exchange
interaction $J$ but also on the local Hubbard interaction of the
conduction electrons $U_c$. This is seen in Fig.~\ref{fig:gaps} where
we plot the gaps measured as the distance between the two maxima for
$U_c = 1$ and $1.5$ vs.\ $J$.  As discussed for the Anderson-Hubbard
model,\cite{Schork97b} our numerical procedure does not allow to
proceed to very low temperatures. As the gaps have not yet completely
converged we only repeat the following qualitative remark which
applied also for the Anderson-Hubbard model: The larger $U_c$, the
larger the resulting gap, and as in the case of the Anderson-Hubbard
model, the quasiparticles of the Hubbard model are responsible for the
screening of the $f$ moment.

\begin{figure*}
\centerline{\psfig{file=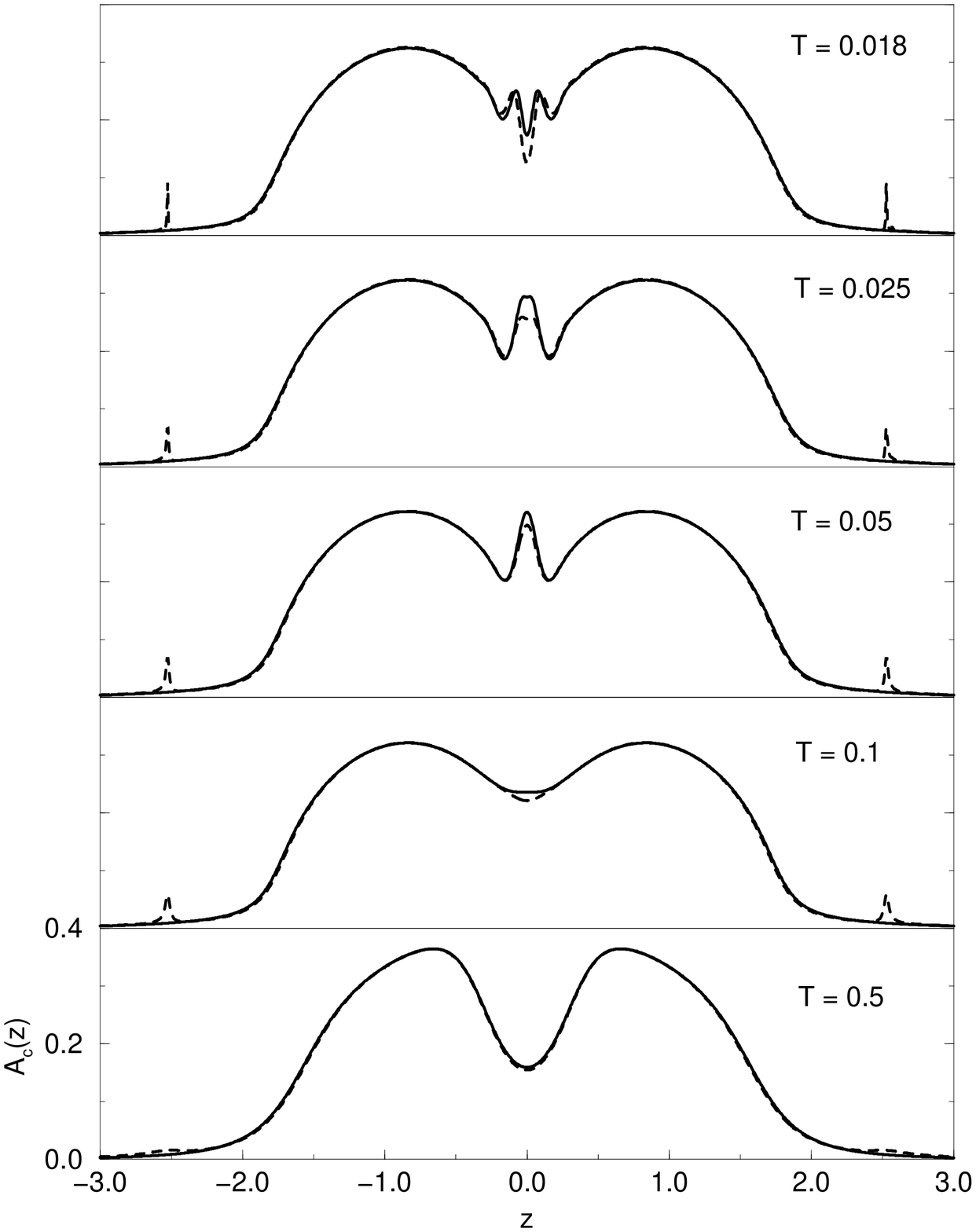,width=8cm}
\psfig{file=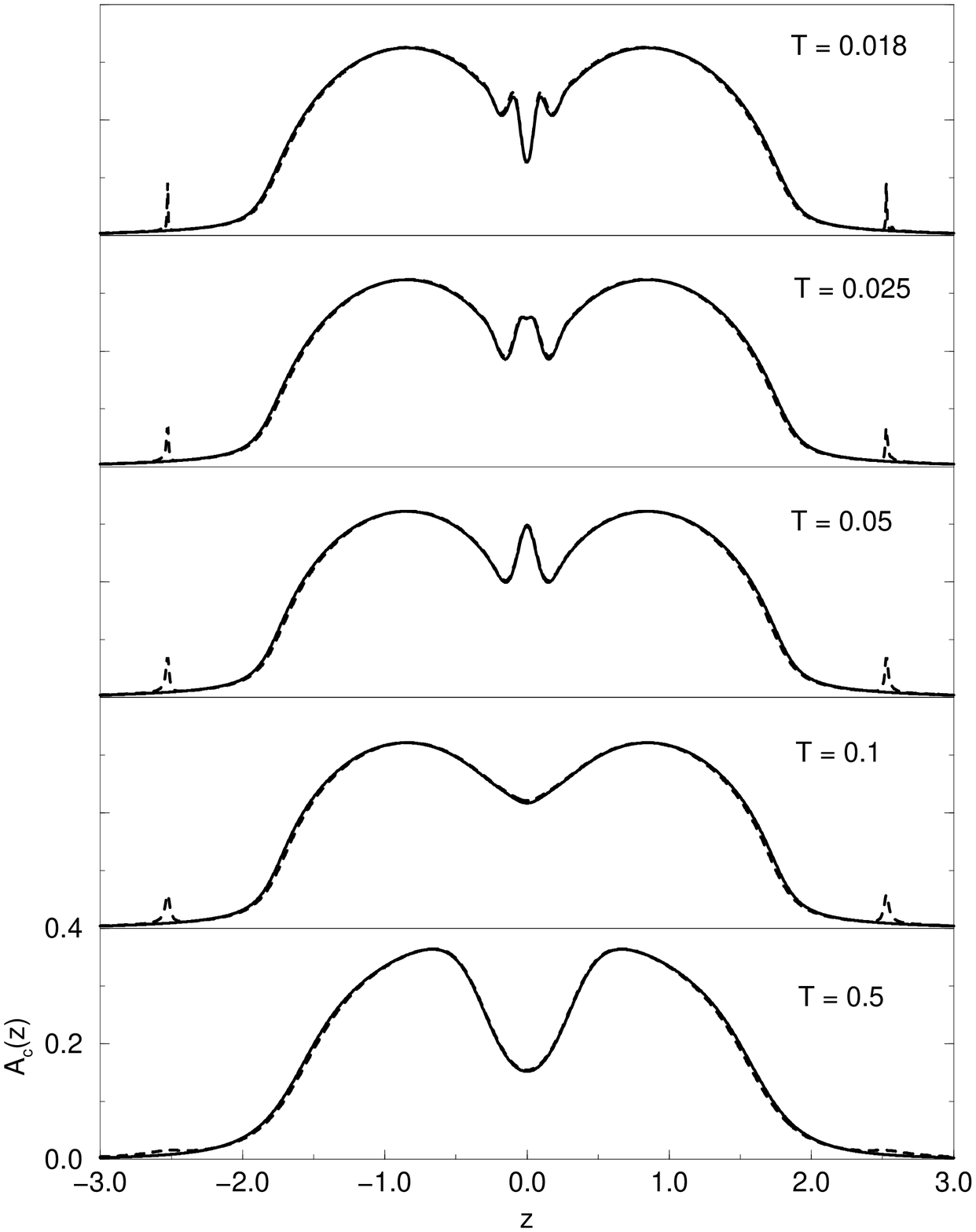,width=8cm}}
\caption{Local density of states $A_c(z)$ in the Kondo-Hubbard model
(solid lines, $J = 0.04851, U_c = 1.479$, left, and $U_c = 1.5$,
right) and the Anderson-Hubbard model (dashed lines, $U_c = 1.5, U_f =
5, V = 0.2$) at different temperatures.}
\label{fig:spec_uc149_j005}
\end{figure*}
Next, we compare these findings to the results of the Anderson-Hubbard
model\cite{Schork97b} in Fig.~\ref{fig:spec_uc149_j005} (left). The
parameters of the Kondo-Hubbard model, $U_c = 1.479$ and $J=0.04851$,
were chosen such that the model originates from an Anderson-Hubbard
model with $U_c = 1.5$, $U_f = 5$, and $V = 0.2$ --- see
Sec.~\ref{sec:mapping}. The most obvious difference is the presence of
the $f$ peak at $z\sim\pm U_f/2$ in the latter case which corresponds
to the possible charge fluctuation on the $f$ orbital. Except for
this, we find that the two spectra are similar. Since the
Kondo-Hubbard model was derived from the Anderson-Hubbard model by
projecting out charge fluctuations on the $f$ orbital this should be
expected for a small hybridization $V$ where these charge fluctuations
are unimportant.

However, close to the chemical potential the two spectra only qualitatively
agree. This should not be too surprising, as in the canonical transformation
the one-particle operators should be transformed as well in order to obtain
equivalent spectra,\cite{Eskes94} see also App.~\ref{app:can}.
Besides higher Green's functions, one finds contributions from ${\cal
G}_{cf}$ and ${\cal G}_f$ of the Anderson-Hubbard model to the $c$
spectral function of the Kondo-Hubbard model. We end this section by
reporting a strange observation. If we ignore the change of $U_c$ when
passing from the Anderson-Hubbard model to the Kondo-Hubbard model, a
difference between the two spectra can hardly be seen at small $V$,
Fig.~\ref{fig:spec_uc149_j005} (right). The rather strong influence of
$U_c$ on the spectra ($U_c$ changed by 1.4\% from
Fig.~\ref{fig:spec_uc149_j005} (left) to
Fig.~\ref{fig:spec_uc149_j005} (right)) is due to the proximity to the
metal-insulator of the bare conduction electrons which in our scheme
occurs at $U_c \sim 1.7$. For larger values of $V$ this finding does
not hold true, but the agreement in total gets worse since charge
fluctuations become more important.


\section{Conclusions}
\label{sec:conclusion}

In summary, we investigated a Kondo lattice model with correlated conduction
electrons within the dynamical mean-field theory. To this end we mapped the
model onto an impurity model in which a unit cell of the lattice is embedded
self-consistently into a bath of free electrons. This impurity model was
studied numerically within the resolvent perturbation theory by applying the
non-crossing approximation.

The parameters of the Kondo-Hubbard model were related to the parameters of
the Anderson-Hubbard model by matching the transition energies within a single
unit cell. It was shown that this is equivalent to performing a canonical
transformation within a single unit cell.

The resulting spectral functions (local density of states) behave similarly as
in the Anderson-Hubbard model: At relatively high temperatures the conduction
electrons are almost decoupled from the $f$ electrons and, as in the pure
Hubbard model, we observe a quasiparticle peak emerging when the temperature
is lowered. When the temperatures is reduced further, the quasiparticle peak
splits and a gap opens. As in the case of free conduction electrons, one might
envisage that a (local) Kondo effect generates resonances at the chemical
potential which hybridize with the quasiparticle states leading to a gap.

The correlations among the conduction electrons, however, do not only show up
in the formation of a quasiparticle peak at high temperatures, but also
influence the size of the gap at low temperatures. In agreement with the
Anderson-Hubbard model, we find that the low-energy scale is increased as the
correlations become stronger.

Both features result from certain transitions in the impurity model.
The first one, which corresponds to the Hubbard quasiparticle peak, is related
to the (local) transition where the $c$ orbital changes from single to double
(empty) occupancy, whereas the second one, leading to the gap, stems from the
singlet-triplet transition of the $c$-$f$ molecule which has a lower energy.
The considered impurity model is a ``minimal'' model which contains two
different excitation energies and the Kondo-Hubbard model is a minimal model
to generate the two features mentioned above. Therefore it deserves continued
interest.

\appendix
\section{Canonical transformation}
\label{app:can}

The traditional way of deriving Kondo models from Anderson models is to
perform a canonical transformation
\begin{equation}
H_{\rm K} = e^{S} H_{\rm A} e^{-S}
\end{equation}
such that there is no $c$-$f$ hybridization in $H_{\rm K}$ and the $f$
occupancy is conserved.\cite{Schrieffer66} Usually this is done in
perturbation theory with respect to the $c$-$f$ hybridization $V$. We will
perform this transformation on the level of the impurity models and rewrite
our impurity model as
\begin{eqnarray}
H_{\rm A,imp} & = & H_0 + H_1 ~, \\
H_0 & = & H_{\alpha} + H_{\alpha c} + H_c + H_f ~, \\
H_1 & = & H_{c f} ~.
\end{eqnarray}
The transformed Hamiltonian is given by
\begin{eqnarray}
H_{\rm K,imp} & = & e^S H_{\rm A,imp} e^{-S} = e^{\cal S} H_{\rm A,imp} 
\label{can_trans}
\\
& = & H_0 + H_1 + {\cal S}_1 H_0 
+ \frac{1}{2} {\cal S}_1^2 H_0 + {\cal S}_1 H_1 
\nonumber \\
&&
+ {\cal S}_3 H_0 + \frac{1}{2} {\cal S}_1^2 H_1 + \frac{1}{6} {\cal S}_1^3 H_0
+ {\cal S}_3 H_1 + {\cal S}_3 {\cal S}_1 H_0 \nonumber \\ 
&& + \frac{1}{6} {\cal S}_1^3 H_1 + \frac{1}{24} {\cal S}_1^4 H_0
+ o(V^5)
\end{eqnarray}
where ${\cal S} A = [S,A]_-$ and ${\cal S}_\nu \sim V^\nu$. Charge
fluctuations on the $f$ orbital are eliminated by choosing $\cal S$ such that
there are no contributions of odd power in $V$. Hence,
\begin{eqnarray}
{\cal S}_1 H_0 = - H_1 & ~\Rightarrow~ & S_1 = \frac{1}{{\cal L}_0} H_1~,
\\
{\cal S}_3 H_0 = - \frac{1}{3} {\cal S}_1^2 H_1  & ~\Rightarrow~ &
S_3 = \frac{1}{3} \frac{1}{{\cal L}_0} {\cal S}_1^2 H_1 ~.
\label{S}
\end{eqnarray}
where the Liouvillean ${\cal L}_0$ is defined by ${\cal L}_0 A = [H_0, A]_-$.
The effective Hamiltonian which conserves the $f$ occupation is then given by
\begin{equation}
H_{\rm K,imp} = H_0 + \frac{1}{2} {\cal S}_1 H_1 + \frac{1}{8} {\cal S}_1^3
H_1 ~.
\end{equation}

In order to reproduce the previous results from Sec.~\ref{sec:mapping} we will
neglect $H_{\alpha c}$ and thus replace ${\cal L}_0$ in Eqs.~(\ref{S}) by
${\cal L}_c + {\cal L}_f$. The Liouvillean is easily inverted by decomposing
$c^\dagger_\sigma = \hat c^\dagger_\sigma + \bar c^\dagger_\sigma$ where
\begin{equation}
\hat c^\dagger_\sigma = c^\dagger_\sigma (1-n^c_{-\sigma}), \qquad
\bar c^\dagger_\sigma = c^\dagger_\sigma n^c_{-\sigma}
\end{equation}
are eigenoperators of ${\cal L}_c$ with eigenvalues $-U_c/2$ and
$U_c/2$, resp.\ (similarly for $f^\dagger_\sigma$). We immediately obtain
$S_1$ 
and find as effective Hamiltonian within the space $n_f = 1$:
\begin{eqnarray}
H_{\rm K,imp} & = &  H_\alpha + H_{\alpha c} + H_c \\
&& + J S_f S_c 
+ \delta U_c \left( -\frac{1}{2} n^c + n^c_\uparrow n^c_\downarrow \right) ~,
\end{eqnarray}
where we dropped a constant and
\begin{eqnarray}
J & = & \frac{8V^2}{U_f+U_c} - \frac{128 V^4}{(U_f+U_c)^3} ~,
\\
\delta U_c & = & - 4V^2 \left( \frac{1}{U_f-U_c} - \frac{1}{U_f+U_c} \right) 
\nonumber \\
&& + 16V^4 \left( \frac{1}{(U_f - U_c)^3} - \frac{4}{(U_f+U_c)^3} \right) ~.
\end{eqnarray}

These are just the results~(\ref{J}) and (\ref{Uc}) found in the previous
section by comparing the excitation energies of the impurity in both models.

Note that we neglected the coupling of the $c$-$f$ molecule to the bath in
this derivation. In reality the $\alpha$ operators should change in the
canonical transformation as well. This approximation corresponds to just
matching the molecular excitation energies in the previous section. Note also
that the canonical transformation was performed on the impurity
model. Therefore no RKKY interaction is generated to forth order in $V$.

When comparing spectral functions of models related by canonical
transformations one should bear in mind that the creation operators should be
transformed as well.\cite{Eskes94} This is seen as follows: The Hamiltonians
in both models are related by (\ref{can_trans}), the ground states via
\begin{equation}
|\Psi_K\rangle = e^S |\Psi_A\rangle ~.
\end{equation}
Thus, the $c$ correlation function, e.g., of the Kondo-Hubbard model is given
by
\begin{eqnarray}
{\cal C}_K(z) 
& = & 
\left\langle \Psi_K \left| c_\sigma^{\phantom{\dagger}} 
\frac{1}{z-H_K} c_\sigma^\dagger \right| \Psi_K \right\rangle 
\nonumber\\
& = & 
\left\langle \Psi_A \left| (e^S)^\dagger c_\sigma^{\phantom{\dagger}} 
(e^{-S})^\dagger \frac{1}{z-H_A} e^{-S} c_\sigma^\dagger e^S 
\right| \Psi_A \right\rangle
\end{eqnarray}
(and correspondingly the Green's function). In particular, the $c$ Green's
function in the Kondo-Hubbard model is not identical to the one in the
Anderson-Hubbard model. Given $S_1$ we find to first order
\begin{eqnarray}
e^{-S} c_\sigma^\dagger e^S & = & c_\sigma^\dagger + 2 V 
\left\{ \frac{2U_c}{U_f^2-U_c^2} \times \right. \nonumber \\ 
&& 
\left( -c^\dagger_\sigma c^\dagger_{-\sigma} f^{\phantom{\dagger}}_{-\sigma}
+c^\dagger_\sigma f^\dagger_{-\sigma} c^{\phantom{\dagger}}_{-\sigma}
+f^\dagger_\sigma c^\dagger_{-\sigma} c^{\phantom{\dagger}}_{-\sigma} \right) 
\nonumber \\
&& \left. -\frac{1}{U_f-U_c} \hat f_\sigma^\dagger 
+\frac{1}{U_f+U_c} \bar f_\sigma^\dagger  \right\} ~.
\end{eqnarray}

\acknowledgements

We would like to thank K.\ Fischer for useful discussions and J.\ Schmalian
for numerical advice. One of the authors (T.S.) would like to thank the Gunma
University Foundation for financial support and the Gunma University at Kiryu
for its kind hospitality.


\end{document}